# Temporary changes in large-scale memory neural networks after fear learning and extinction in healthy adults

*Kirill V. Efimov[1,2], Alina O. Temereva[2], Aleksey M. Ivanitskiy[2], Sergey I. Kartashov[3], Olga V. Martynova[1,2]*

[1]*Institute of Cognitive Neurosciences, National Research University "Higher School of Economics"*
[2]*Institute of Higher Nervous Activity and Neurophysiology of the Russian Academy of Sciences*
[3]*National Research Institute "Kurchatov Institute"*

ABSTRACT

The analysis of the functional connectivity of brain sections associated with fear-conditioned training is one of the methods of study of memory neural networks. Before, the majority of studies focused on the functional connectivity of the brain regions that are data-driven for emotional education such as amygdalae and areas of the ventromedial prefrontal cortex in the resting state right after the formation of a fear-conditioned reflex. In the present study, the authors applied the methods of the theory of graphs to search for changes in the functional connectivity at the level of the brain in the resting state in a week's dynamics after the extinction of a conditioned reflex with partial reinforcement.

The authors compared the data on the functional connectivity of 23 healthy volunteers from the experimental group and 19 volunteers from the control group. The experimental group underwent 5 functional magnetic resonance imaging sessions: before fear-conditioned training (formation of a conditioned reflex with partial reinforcement), during conditioned reflex extinction, a day and a week after the extinction of a conditioned reflex. Volunteers from the control group underwent 3 functional magnetic resonance imaging sessions: the second and third were performed a day and a week after the first session, respectively. The authors conducted a statistical analysis of the parameters of functional connectivity graphs between 246 functional areas of the brain according to the Brainnetome Atlas. The authors calculated the parameters for the nodes of all the areas, identified specific subnetworks activated in conditioned reflex extinction sessions in the experimental group, and monitored their changes in time in comparison with the functional connectivity dynamics of networks in the control group.

The most significant changes were observed in the functional connectivity of the left parahippocampal area. In particular, the rostral part of the left parahippocampal gyrus became the center of a new subnetwork connected with the rostral part of the left hippocampus in all the sessions and after the extinction of a conditioned reflex and the lateral part of the left amygdala right after the extinction and a day after the extinction of a conditioned reflex. A week after the extinction of a conditioned reflex, the rostral part of the left parahippocampal gyrus also had more connections with the areas of the middle frontal gyrus in comparison with the parameters of the baseline resting state to the stimulus. Besides, the posterior part of the left parahippocampal gyrus during the extinction and after the extinction of a conditioned reflex lost the connection with the rostral hippocampus and preserved the connection only with the left caudal hippocampus.

Earlier, elevated activation of the left parahippocampal gyrus (Alvarez *et al*., 2008) and elevated activation in the left posterior parahippocampal gyrus during the training with fear conditioning (Koelsch, 2017) were observed. The obtained data showed a similar but more detailed transformation of functional connectivity in this area. Besides, these changes remained within one week from the moment of the extinction of a conditioned reflex, which could be explained by the chosen paradigm of conditioned reflex with partial reinforcement that led to a slower extinction than a paradigm with 100% reinforcement.

LIST OF ABBREVIATIONS

fMRI – functional magnetic resonance imaging

FC – functional connectivity

CR – conditioned reflex

BOLD – blood-oxygen-level dependent signal

CMFT – constant magnetic field of the tomograph

RF – resonance frequency

NMR – nuclear magnetic resonance

ROI – region of interest

PTSD – posttraumatic stress disorder

RS_0 – resting state before fear-conditioning training

FE – session during conditioned reflex extinction

RS_FE – session after the extinction of a conditioned reflex

RS_1 – session a day after the extinction of a conditioned reflex

RS_7 – session a week after the extinction of a conditioned reflex

Control_1 – session in the control group

Control_2 – session in the control group a day after the first session

Control_3 – session in the control group a week after the first session

NBS – Network-Based Statistic

MNI – Montreal Neurological Institute

DMN – Default Mode Network

PCC – paired correlation coefficient

ParCC – partial correlation coefficient

FDR – false discovery rate

Amyg – corpus amygdaloideum

Hipp – hippocampus

PhG – parahippocampal gyrus

vmPFC – ventromedial prefrontal cortex

INS – insular lobe

CG – callosal gyrus

BG – basal ganglia

Tha – thalamus

SFG – superior frontal gyrus

MFG – middle frontal gyrus

IFG – inferior frontal gyrus

OrG – orbital gyrus

PrG – precentral gyrus

STG – superior temporal gyrus

MTG – middle temporal gyrus

ITG – inferior temporal gyrus

pSTS – superior temporal sulcus

SPL – superior parietal lobule

IPL – inferior parietal lobule

PCun – precuneus

PoG – postcentral gyrus

MVOcC – medioventral occipital cortex

LOcC – lateral occipital complex

FuG – fusiform gyrus

FC – fear conditioning

CS+ – conditioned stimuli +

CS- – conditioned stimuli -

US – unconditioned stimuli

GSR – galvanic skin response

ICA – independent component analysis

FWER – family-wise error rate

ANOVA – analysis of variance

INTRODUCTION

Training with the formation of a fear-conditioned reflex (CR) is one of the main methods widely used for the identification of brain sections involved in the training of avoidance of negative effects (LaBar *et al*., 1998; Harnett *et al*., 2016). The application of this method of study on animals showed that the data-driven brain areas associated with memory on negative stimuli and the training of their avoidance were hippocampus (Phillips *et al*., 1992; Lorenzini *et al*., 1996; Sacchetti *et al*., 2002; Maren, 2008), amygdalae (Phillips *et al*., 1992; Royer *et al*., 2002; Samson *et al*., 2005; Maren, 2008; Pape *et al*., 2010), prelimbic cortex (Sotres-Bayon *et al*., 2012), infralimbic cortex (Do-Monte *et al*., 2015).

The studies on brain mapping with functional magnetic resonance imaging (fMRI) also confirm that corpus amygdaloideum (Amyg), hippocampus (Hipp), parahippocampal gyri (PhG), ventromedial prefrontal cortex (vmPFC), insular cortex (INS), and anterior callosal gyri (ACG) take part in the processing of the information, training to avoid, and long-term storage of the memory on negative stimuli (meta-analysis by Fullana *et al*., 2016, 2018). A separate direction of studies in this area on humans is determined the method of analysis of functional connectivity (FC) between the areas of the brain before and after negative stimuli (Dopfel *et al*., 2018; Guadagno *et al*., 2018) and the analysis of changes in FC in clinical groups with different disorders that involve functioning of memory neural networks and cognitive control such as posttraumatic stress disorder (PTSD) (Francati *et al*., 2007; Hughes, 2011; Zhu *et al*., 2017), anxiety disorders (Nakao *et al*., 2011; Mochcovitch *et al*., 2011), and depression (Kerestes *et al*.; 2014; Zhu *et al*., 2017; Ambrosi *et al*., 2017).

The majority of studies in this direction are focused on the FC of Amyg, the data-driven region in the processing of emotions and emotional education (Schultz *et al*., 2012), the whole brain (Roy *et al*., 2009; Kerestes *et al*., 2017), and regions of the brain cortex that control the process of training and memory in cases of negative stimuli. Among them, the closest attention was paid to the associations between Amyg and vmPFC (Schultz *et al*., 2012; Guadagno *et al*., 2018), ACG (Schultz *et al*., 2012; Greco, 2016), INS (Jenkins *et al*., 2017), Hipp (Williams *et al*., 2001; Greco, 2016). The analysis of FC was performed by different methods. One of the hypothesis-driven approaches is the method of FC revision between the regions of interest (ROI), when the connection between the data-driven regions are studied that showed their involvement in the process (for example, the connection between Amyg and vmPFC (Schultz *et al*., 2012; Guadagno *et al*., 2018). Another hypothesis-driven approach is the analysis of the connection between the data-driven region and the rest of the brain (seed-based approach), when the connection between Amyg and fMRI signal from all brain voxels is studied (Biswal *et al*., 1997; Cordes *et al*., 2000; Jiang *et al*., 2004). Finally, for the analysis of FC of large-scale neural networks without the associations with the data-driven areas or connectomes, the methods of the theory of graphs are used, especially, for the comparison of control and clinical groups (Salvador *et al*., 2005; Liu *et al*., 2008; Wang *et al*., 2010). The latter provides a description of networks and nodes from the point of view of all brain regions. Usually, the ROI according to the functional atlas are the graph nodes (vertices) and the presence of significant connectivity between two ROI is the edge between these nodes that form a network of FC. Further, some studies revealed consistencies in the transformation of networks of FC of the whole brain and the changes in the integration of separate ROI, and in the structure of connections and paths (sequence of connections) between certain ROI in the chosen area or the whole brain (Bullmore & Sporns, 2009; Wang *et al*., 2010; Power *et al*., 2011). (Graphs are better than separate regions).

It should be highlighted that evidence was obtained that functional transformations determining recollections on the events could be revealed with the FC method not only during the time of stimulus but also for some time at the resting time after the negative stimulus (Feng *et al*., 2015; Schultz *et al*., 2012; Hermans *et al*., 2017; Martynova *et al*., 2020). These data serve as an indirect confirmation of the theory of long-term neuroplastic transformations at the level of micro and macro networks during the formation of memory and training for the avoidance of negative stimuli verified for animals (Alberini, 2009; Liang *et al*., 2014). Thus, recent studies performed by the authors showed that the transformation in the FC of the left and right amygdalae could asymmetrically fluctuate in the parameters of fMRI in the resting condition in healthy volunteers registered with a one-week interval (Tetereva *et al*., 2019). Further comparison of FC in the right and left amygdalae in the control group that did not receive negative stimuli and the group of volunteers that underwent the procedure of fear conditioning and CR extinction with partial negative reinforcement demonstrated a residual increase in the FC of the left amygdala with the prefrontal cortex in the experimental group in the resting condition even a week later (Martynova *et al*., 2020).

In previous studies, the authors focused on the dynamics of FC of the amygdalae using the seed-based method. In the present study, the authors aimed to investigate possible large-scale changes in the FC at the level of the whole brain after the formation and extinction of CR on negative stimuli in healthy volunteers. Thus, the authors intended to check a hypothesis that during the formation of a CR, long-term neuroplastic transformations occurred in the connections of large-scale neural networks involved in the formation and memorizing of the stimulus that had an emotional factor and that these transformations could be seen in the FC of the brain areas in the resting state after the stimulus. The authors suggested that such long-term transformations in large-scale networks remained in cases with a CR formation with partial reinforcement that led to a slower extinction than the paradigm with 100% reinforcement due to a high indeterminacy (Milad *et al*., 2007; Schiller *et al*., 2013; Feng *et al*., 2015). To check this hypothesis, the authors evaluated brain connectomes in the resting state using the theory of graphs. To reduce the loss of connections due to a correction for multiple comparisons, for the calculation of the parameters of graphs, the authors used 246 brain areas averaged by the FC, taken from the Brainnetome Atlas (Fan *et al*., 2016) that accounts for anatomic and functional brain parcellation based on the BrainMap database of the brain activation (Laird *et al*., 2009). Just like in previous studies, the authors compared the dynamics of functional connectomes in the control (no stimuli) and experimental (fear conditioning and CR extinction) groups starting from the baseline (initial) fMRI of the parameters of brain activity in the resting state including a day and a week after the first session. For the statistical check of the measurements in the connectomes, the authors used the permutation test NBS (network-based statistics) (Zalesky *et al*., 2010) paying special attention to the most significant changes in the subnetworks in the resting condition in the areas involved in the emotional training and formation of emotional memory: Amyg, BG (basal ganglia), INS, ACG, PFC (prefrontal cortex), Tha (thalamus), MTG (middle temporal gyrus), Hipp, and PhG (Tyng *et al.*, 2017). The obtained results showed that the most consecutive transformations were observed in the hippocampal area.

MATERIALS AND METHODS

*Participants*

The experiment included 42 healthy volunteers. The participants were divided into experimental and control groups. The experimental group included 23 volunteers (15 men and 8 women) aged 18–30 years old (23.90 ± 3.93 years old). The control group included 19 volunteers (14 men and 5 women) aged 18–32 years old (26.20 ± 3.89 years old). All the participants were right-handed with normal eye-sight. Among the participants, there were no subjects that had brain injuries in the anamnesis, contraindications to magnetic resonance imaging, received drugs affecting the central nervous system, had neurologic and psychic disorders in the anamnesis, drug addicts, alcohol abusers, and smokers, pregnant women, and subjects older than 32 years old. The protocol of the experiment was approved by the Ethical Committee of the Institute of Higher Neural Activity and Neurophysiology of the Russian Academy of Sciences according to the requirements of the Helsinki Declaration. All the volunteers signed a form of informed consent before the experiment.

*Experimental procedure*

The experimental group underwent fMRI of the brain in the resting state before the formation of a CR with negative reinforcement (RS_0), during the procedure of CR extinction (E), right after the extinction (RS_FE), a day (RS_1), and a week (RS_7) after the extinction.

The control group underwent fMRI in the resting state three times: second and third sessions were performed with a one-day and one-week interval after the first session, respectively.

The duration of each session was 10 minutes. Between the sessions, there were 5–10-minute breaks depending on the requirements for the participant, technical adjustment, and collection of structural magnetic resonance imaging. During the session in the resting state, the participants were instructed to keep calm, lie with shut eyes, not to move, and not to concentrate on certain thoughts. The heads of the participants were stabilized with soft barriers to reduce the artifacts from movements.

Fear-conditioning consisted of the presentation of two consecutive symbols of three types: one neutral and two conditioned stimuli + (CS+) (30% and 70% in the first sequence and 70% and 30% in the second sequence). The CS was weak current set for a pain response of a participant. The time of CS was 500 μs that was initiated right after the presentation of a symbol on a white screen. Before the presentation of a symbol, a white cross was shown on the screen for 2 seconds. The duration of the presentation of symbols reinforced by CS was 4–8 seconds (randomly) with a 2-second interval. After the presentation of a symbol with possible CS, participants were shown the white screen for 8–12 seconds (randomly) with a 2-second interval. The general time of the session with fear-conditioning was 8 minutes 54 seconds.

During the session of extinction of a CR (FE), the participants were offered the same symbols in another pseudo-random sequence and without fear-conditioning for 10 minutes. The patients were instructed to wait for negative stimuli.

The paradigm of the experiment and the data coincide with those described in the earlier publications (Tetereva *et al*., 2019; Martynova *et al*., 2020). A complete description is presented in the Annex (Annex, Figure 1).

The control of training with fear conditioning was performed by the registration of galvanic skin response (GSR) that used direct current. The voltage was 0.9 V with a frequency of 4000 Hz. Ag/AgCl electrodes (Medical Computer Systems LTD, Moscow, RF) with electrode gel were placed on the left palm with a 14-mm clearance. A digitalized signal was sent via a frequency window with a maximum value of 16 Hz and purified from artifacts of movements using BrainVision Analyzer 2.1 (Brain Products, GmbH, Germany). GSR for each session was calculated by the deduction of the skin resistance measured 2 seconds before the presentation of a symbol from the maximum skin resistance during the interval with negative stimuli. The values of GSR were normalized for each participant and compared using ANOVA for repeated measurements. Further, a post hoc comparison between each CS+ and CS– in both sessions of training with fear conditioning and sessions with the extinction of a CR was performed.

*Data collection*

Data collection was performed using a 3T scanner Magnetom Verio Siemens (Germany) equipped with a 32-canal head coil. For each participant, sagittal images of anatomical structures of the brain were obtained using gradient sequence T1 MP-RAGE: TR 1470 ms, TE 1.76 ms, FA 9°, 176 1 mm sections with a 0.5-mm clearance, 320 mm field of vision with 320×320 matrix. Further, functional images were obtained using a T2-weighted echo-planar sequence with the following parameters: TR 2000 ms, TE 20 ms, FA 90°, 42 2-mm-thick sections with a 0.6-mm clearance, the field of vision 200 mm, 98×98 matrix. To reduce the spatial distortion of the magnetic field, the authors obtained amplitude and phase images using the algorithm "FieldMap" with the following parameters: TR 468 ms, TE1 4.92 ms, TE2 7.38 ms, FOV 200 mm, 42 section, FA 60°.

*fMRI data pre-processing*

Pre-processing of the data was performed using MELODIC, a part of the software package FSL (FMRIB's Software Library, www.fmrib.ox.ac.uk/fsl). During the pre-processing of the data, the authors performed the following procedures: correction for movements (MCFLIRT), slice-timing correction with a phase shift of time sequences in Fourier space, removal of sections that did not belong to the brain, spatial smoothing using the Gaussian nucleus with full width at half maximum 5 mm, normalization of the mean intensity of the whole 4D-massive using a unified multiplier, high-pass temporal filtration with the section frequency of 0.01 Hz (Jenkinson *et al.*, 2002; Smith, 2002).

B0-distortions were removed using the algorithm of unfolding B0. The registration of functional images in the individual anatomic and standard spaces MNI152 2 mm$^3$ was performed using FLIRT (Jenkinson *et al.*, 2001, 2002). Further, the authors performed independent component analysis (ICA) using MELODIC (in 3.14). For each signal of each participant, 38 independent components were obtained (Hyvärinen, 1999; Beckmann and Smith, 2004). Additional denoising was performed using FIX 1.068 (FMRIB's ICA-based Xnoiseifier (Salimi-Khorshidi *et al.*, 2014; Griffanti *et al.*, 2014). Further, component-dependent removal of artifacts of movement (ICA-AROMA) was carried out (Pruim *et al.*, 2015a, 2015b). Firstly, AROMA was applied to 15 individual sessions (per 5 randomly chosen from the RS_0, FE, and RS_FE groups) in the classification mode to identify the components that depended on movement. After that, the results were visually checked according to the guidelines (Griffanti *et al.*, 2017) for the identification of additional artifact-dependent components including fluctuations of cerebrospinal fluid in the brain ventricles. Further, FIX was trained on preliminary classified 15 individual sessions and automatic classification was applied to the rest individual sessions. Noisy components were removed using FIX in the mode of movement distortion removal (24 repressors). Further, the obtained data was passed through a frequency filter in the range of 0.01–0.1 Hz using the algorithm "3dTproject" AFNI (Cox, 1996).

*Statistical analysis*

For time sequences of the values of the BOLD signal in the ROI, the authors calculated the partial correlation coefficient (ParCC) for the BOLD signal (Marrelec *et al.*, 2006). Fisher z-transformation was used to obtain normal distribution for each ParCC for participants that were applied one-sample Student's t-test. The null hypothesis is that the mean ParCC equals zero. An alternative hypothesis is that the mean ParCC significantly differs from zero. The results were adjusted for Benjamini-Hochberg multiple comparisons ($p < 0.05$). As a result, significant ParCC between the ROI were obtained for each session.

The sessions were compared using the NBS method (Zalesky *et al.*, 2010) that involved a general linear model together with the permutation test and controlled family-wise error rate (FWER) during the comparison by each connection for two and more networks. For the evaluation of the influence of intrasession factors, the authors used the method of factorial analysis one-way ANOVA with 5 levels of sessions for the experimental group and 3 levels for the control group. For the evaluation of the influence of day and group factors, the authors used the method of factorial analysis two-way ANOVA with 2 levels for the group factor (control and experimental) and 3 levels for the day factor (1 day, 2 days, 7 days). Further, post hoc Tukey's test was used (the procedure of multiple comparisons with the error control depending on the number of comparisons for differences in certain sessions within a group and between the groups).

*Graph theory approach*

Connections that passed the statistical threshold were presented as a graph for the respective session in Gephi 0.9.2 software.

For each graph as a whole, the authors calculated the mean clustering coefficient, average shortest path length, and global efficiency. The clustering coefficient reflects the elasticity of the network nearby a certain node, and the mean clustering coefficient reflects the variability of the network as a whole. The average shortest path length and global efficiency describe the capacity of a network for parallel data transmission.

For the comparison of the FC networks, the small-world coefficient is widely used, which is the ratio of the mean clustering coefficient to the mean shortest path length standardized for the same parameters of a random graph (Watts and Strogatz, 1998). The small-world coefficient is a parameter of the computational efficiency of a network (Achard and Bullmore, 2007; Bullmore and Sporns, 2009; Rubinov and Sporns, 2010, 2011) and is used for networks, in which the average shortest path length is significantly lower than the number of nodes (Milgram, 1967).

Besides, the authors calculated node eigenvector centrality (relative degree of nodes adjoining to the given) and closeness centrality (distance from the given node to the rest nodes). Each parameter calculated by the individual connectomes of the participant underwent ANOVA analysis for the identification of statistical differences between the sessions.

## RESULTS

*Network parameters*

The average shortest path length demonstrated an increase in sessions with CS+. The mean coefficient of node clustering and global efficiency demonstrated a decrease in sessions with CS+. All parameters returned to the baseline state with time. Complete connectomes for each session are described in the Annex (Annex, Table 1.1–1.8, Figure 2.1–2.8).

## Table 1. Global network parameters

|  | RS_0 | FE | RS_FE | RS_1 | RS_7 | Control_1 | Control_2 | Control_3 |
|---|---|---|---|---|---|---|---|---|
| N of nodes | 245 | 245 | 246 | 246 | 246 | 245 | 245 | 246 |
| N of associations | 510 | 449 | 479 | 481 | 512 | 397 | 408 | 412 |
| Global efficiency | 0.227 | 0.204 | 0.207 | 0.219 | 0.224 | 0.189 | 0.190 | 0.195 |
| Average shortest path length | 5.14 | 5.84 | 5.78 | 5.36 | 5.22 | 6.29 | 6.29 | 6.09 |
| Mean clustering coefficient | 0.17 | 0.15 | 0.14 | 0.13 | 0.18 | 0.12 | 0.12 | 0.08 |
| Small-world coefficient | 10.46 | 12.00 | 9.46 | 9.65 | 11.49 | 11.59 | 10.30 | 7.01 |

## Table 2. Mean parameters of individual connectomes

|  |  | RS_0 | FE | RS_FE | RS_1 | RS_7 | Control_1 | Control_2 | Control_3 |
|---|---|---|---|---|---|---|---|---|---|
| Global efficiency | mean | 0.130 | 0.103 | 0.117 | 0.118 | 0.124 | 0.083 | 0.092 | 0.095 |
|  | SD | 0.007 | 0.009 | 0.011 | 0.009 | 0.013 | 0.013 | 0.014 | 0.009 |
| Average shortest path length | mean | 7.52 | 9.07 | 8.69 | 8.30 | 7.81 | 11.2 | 10.1 | 9.52 |
|  | SD | 0.40 | 0.65 | 1.00 | 0.56 | 0.48 | 1.73 | 1.65 | 0.97 |
| Mean clustering coefficient | mean | 0.0424 | 0.0426 | 0.0500 | 0.0450 | 0.0411 | 0.0394 | 0.0298 | 0.0328 |
|  | SD | 0.0142 | 0.0154 | 0.0172 | 0.0204 | 0.0174 | 0.0164 | 0.0180 | 0.0144 |
| Small-world coefficient | mean | 3.79 | 4.45 | 5.49 | 4.67 | 4.03 | 5.05 | 3.37 | 3.73 |
|  | SD | 1.24 | 1.73 | 1.96 | 1.58 | 1.50 | 2.10 | 1.88 | 1.58 |

## Table 3. Results of the factor analysis by individual connectomes

|  | Intragroup comparison of sessions | | | Intergroup comparison of sessions | | |
|---|---|---|---|---|---|---|
|  | Group | F-value | P-value | Factor | F-value | P-value |
| Global efficiency | Exp | 23.69 | 3.8e-14 | Group | 252.855 | < 2e-16 |
|  | Control | 5.288 | 0.00799 | Day | 0.657 | 0.419 |
| Average shortest path length | Exp | 21.39 | 4.54e-13 | Group | 133.29 | < 2e-16 |
|  | Control | 6.031 | 0.00432 | Day | 5.48 | 0.0208 |
| Mean clustering coefficient | Exp | 0.952 | 0.437 | Group | 8.339 | 0.00459 |
|  | Control | 1.726 | 0.188 | Day | 0.991 | 0.32153 |

| Small-world coefficient | Exp | 3.785 | 0.00637 | Group | 0.131 | 0.718 |
| | Control | 4.277 | 0.0189 | Day | 1.571 | 0.212 |

For the evaluation of parametric differences of individual connectomes between sessions, the authors chose global efficiency, the mean coefficient of clustering, the average shortest path length, and the small-world coefficient that were used for one-way ANOVA for the evaluation of the intragroup sessions, two-way ANOVA for the evaluation of the effects of group and day, and Tukey's test for the comparison of individual sessions within post hoc analysis.

The mean coefficient of clustering did not show significant differences between sessions by the results of Tuckey's tests and ANOVA except for the differences of the groups.

By the results of Tuckey's tests, the coefficient of global efficiency significantly differed between FE and every other session in the experimental group, as well as between RS_0 and RS_FE, RS_0, and RS_1. Besides, by the results of Tuckey's tests, Sessions 1 and 3 significantly differed in the control group. This explains the significance of differences identified by one-way ANOVA. At the same time, two-way ANOVA revealed the group effect and did not reveal the day effect.

By the results of Tuckey's tests, the average shortest path length significantly differed in the comparison of all the sessions in the experimental group with RS_0 except for RS_7, comparison of FE with RS_7, FE with RS_1, RS_FE with RS_7, and Sessions 1 and 3 in the control group. Two-way ANOVA revealed differences in the effects of group and day. Tuckey's test revealed a significant group effect and did not reveal significant differences between the days, although the differences between Days 1 and 3 were on the border of significance ($p = 0.054 > 0.05$).

The small-world coefficient showed significant differences by the results of Tukey's test between the sessions of the experimental group RS_FE and RS_0, RS_FE and RS_7, and Sessions 1 and 2 in the control group. Besides, a significant difference between the sessions within groups was revealed by one-way ANOVA. Neither two-way ANOVA no Tuckey's test revealed significant group and day effects. Two-way ANOVA revealed that the group effect was more significant than the day effect. The day effect was less significant and passed the level of significance only in comparison to the average shortest path length. In general, the differences in the parameters between the groups of participants were more significant than between the days of scanning. The most significant dynamic in the changes of individual connectomes was observed in the average shortest path length and global efficiency. Still, even though the global efficiency coefficient and the average shortest path length in a graph are mathematically associated parameters, there were no significant differences revealed in the global efficiency coefficient between the days of scanning.

*Changes in the parameters of the main connectome hubs*

Below, there are parameters of nodes in complete connectomes assorted by the degree, cut by the parameter of the degree of a node = 6. Complete tables for all 246 nodes of each connectome are presented in the Annex (Annex, Table 2.1–2.8).

**Table 3.1. – 3.5. Parameters of the main connectome hubs**

In the control groups, the main hubs were IFG (inferior frontal gyrus), ITG (inferior temporal gyrus), MFG (middle frontal gyrus), MTG, STG (superior temporal gyrus), SFG (superior frontal gyrus), PoG (postcentral gyrus), PhG, PrG (precentral gyrus), CG (callosal gyrus), MVOcC (medioventral occipital cortex), IPL (inferior parietal lobule), and INS (Annex, Tables 3.1–3.3). In cases with higher degrees, they showed the highest coefficients of centrality in all the groups. Tha formed its cluster.

Compared to the groups of control, RS_0 (Table 3.1) was characterized by an enhanced LOcC (lateral occipital complex), SPL (superior parietal lobule), and PhG, and weak PoG. In general, the distribution of nodes RS_0 was similar to the control groups.

In FE (Table 3.2), IPL_R_6_1 became the main connectome hub. The centrality of the precuneus (Pcun) and orbital gyrus (OrG) increased. OrG was one of the main hubs of the network. The Tha cluster enhanced slightly. PhG and Hipp showed a high coefficient of clustering. Hipp and amygdalae acquired higher eigen centrality, which indicated a higher summed degree of neighboring nodes, even though they were characterized by low closeness centrality like in control groups.

In RS_FE (Table 3.3), IPL remained one of the main network hubs, although the main regions became other areas: IPL – IPL_R_6_4 and IPL_R_6_5. Besides, PrG and Tha became strong hubs. The Tha cluster became a region of the network with the highest degree of neighbors. The second such region was IPL. In Pcun and PrG, the summed degree of neighbors also increased. The highest closeness centrality was observed in IPL, OrG, and Pcun. Slightly lower but relatively high closeness centrality was registered in the amygdala. The areas of the Hipp were characterized by the highest degree of clustering. ITG, MTG, and STG became weaker.

In RS_1 (Table 3.4), a new area IPL – IPL_R_6_2 became the main network hub and acquired the highest degree of neighbors. For the first time, the left IPL became a strong hub. Besides, more PrG areas became stronger hubs. In MVOcC, it was observed less significantly. The highest centrality was observed in LOcC and the main hubs. OrG, amygdala, and the Tha cluster lost their high parameters of centrality.

In RS_7 (Table 3.5), OrG was no longer one of the main network hubs. LocC and Pcun lost high levels of centrality. In STG and ITG, the parameters of the centrality restored to the levels of the control groups, and in MTG, they remained low.

*Dynamics in the changes in the networks in the experimental group*

The results of the NBS test revealed subnetworks that were significantly different between the sessions. Complete results are presented in the Annex (Annex, Figures 2.1–2.2).

Below, there are connections that significantly differed between RS_0 and the rest networks from the experimental group: FE – RS_0, RS_FE – RS_0, RS_1 – RS_0, RS_7 – RS_0, excluding the associations that were registered in Control_1 – RS_0.

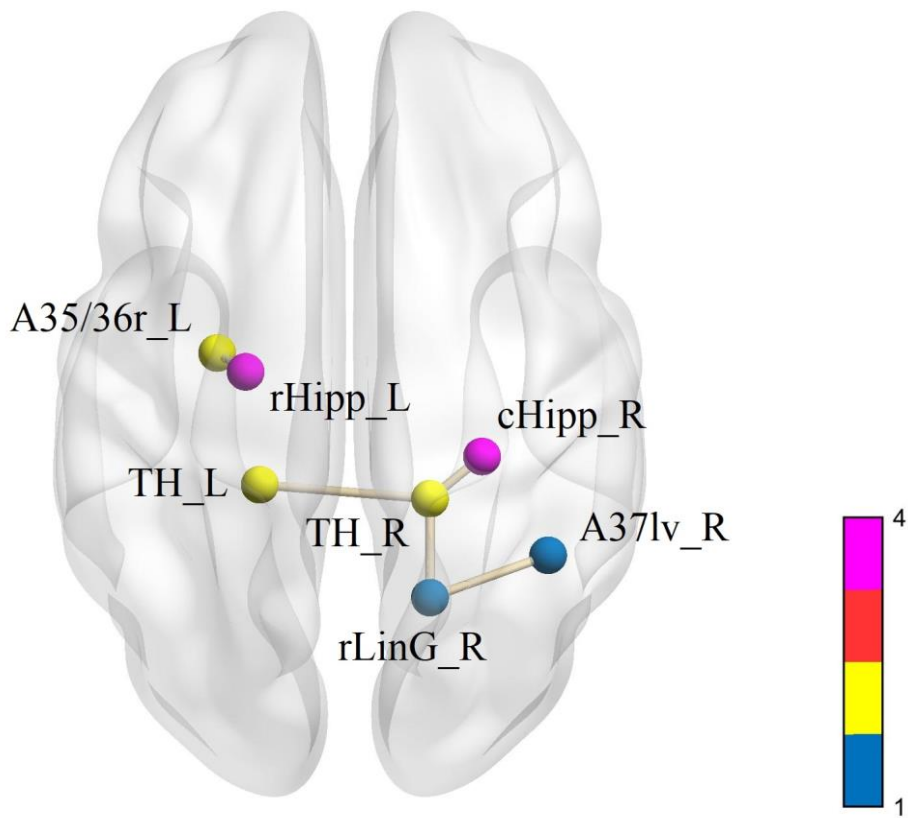

Figure 6. FE – RS_0

Violet – hippocampus

Red – corpus amygdaloideum

Yellow – parahippocampal gyrus

Blue – other areas

In FE, the rostral part of the left Hipp (rHipp_L) formed a connection with the rostral part of the left part of PhG (A35/36r_L) that was not observed in RS_0 (Figure 6). The caudal part of the right Hipp (cHipp_R) was included in a new small subnetwork with the center in the area of TH of the right posterior medial PhG (TH_R). This subnetwork also included the area of TH of the left posterior medial PhG (TH_L), right rostral lingual gyrus (rLinG_R), and lateroventral area 37 of the right FuG (fusiform gyrus) (A37lv_R).

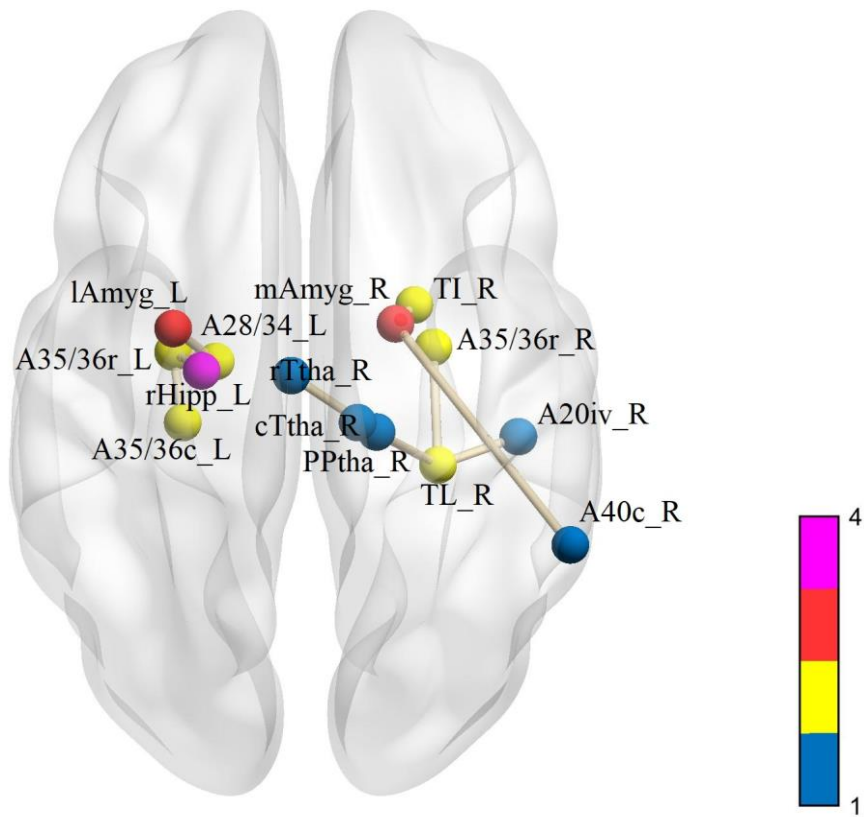

Figure 8. RS_FE – RS_0

Violet – hippocampus

Red – corpus amygdaloideum

Yellow – parahippocampal gyrus

Blue – other areas

In RS_FE, the rostral part of the left PhG (A35/36r_L) became the central hub of a new subnetwork, along with the rostral part of the left Hipp (rHipp_L) connected with the caudal part of the left PhG (A35/36c_L) and area 28/34 of the left entorhinal cortex (A28/34_L), which, in turn, was connected with the left lateral amygdala (lAmyg_L) (Figure 8). Besides, a network appeared that consisted of the right posterior lateral PhG (TL_R), the rostral part of the right PhG (A35/36r_R), the intermediate ventral area 20 of the ITG (A20iv_R), the IPL part of the right Tha (PPtha_R), the caudal part of the right temporal thalamus (cTtha_R), and the rostral part of the right temporal thalamus (rTtha_R). Besides, the medial part of the right amygdala (mAmyg_R) was connected with the area TI of the right temporal agranular INS (TI_R) and caudal area 40 of the right IPL (A40c_R).

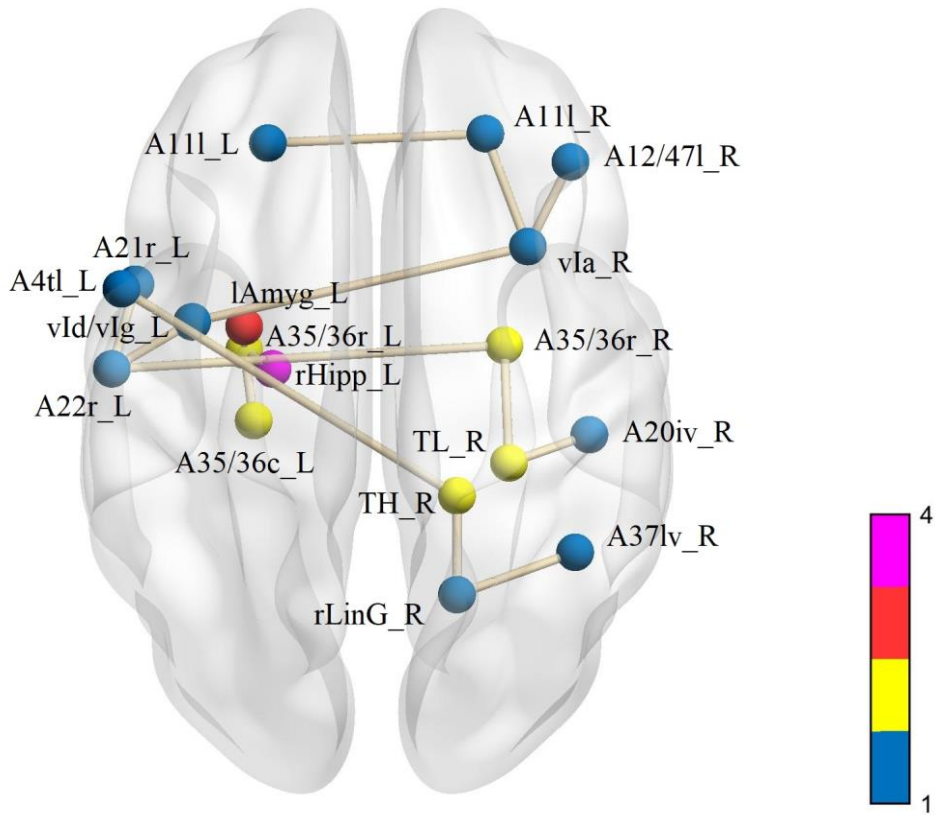

Figure 10. RS_1 – RS_0

Violet – hippocampus

Red – corpus amygdaloideum

Yellow – parahippocampal gyrus

Blue – other areas

In RS_1, the rostral part of the left PhG (A35/36r_L) was connected directly with the left lateral amygdala (lAmyg_L), remained in the center of a new subnetwork (in comparison with RS_0), and included the rostral part of the left Hipp (rHipp_L) and the caudal part of the left PhG (A35/36c_L) (Figure 10). The area 28/34 of the left entorhinal cortex (A28/34_L) was excluded from this new subnetwork. Besides, a new subnetwork expanded that included the right posterior lateral PhG (TL_R), the rostral part of the right PhG (A35/36r_R), and the intermediate ventral region 20 of the ITG (A20iv_R), but instead of the connection between the right posterior lateral PhG (TL_R) and the areas of Tha in RS_FE, the rostral part of the right PhG (A35/36r_R) via the rostral area 22 of the left STG (A22r_L) was connected with the area of a subnetwork that consisted of the left ventral agranular/granular INS (vId/vIg_L), the rostral area 21 of the left MTG (A21r_L), the right ventral agranular INS (vIa_R), the lateral area 12/47 of the right OrG (A12/47l_R), and the lateral areas 11 of the left and right OrG (A11l_L, A11l_R). The connection between the rostral area 22 of the left STG (A22r_L) and left ventral agranular/granular INS (vId/vIg_L) was a new one starting from FE session. Besides, a small network reappeared that consisted of the TH area of the right posterior medial PhG (TH_R), right rostral lingual gyrus (rLinG_R), and the lateroventral area 37 of the right FuG (A37lv_R). This subnetwork, formed in FE together with the caudal part of the right Hipp (cHipp_R) and the TH area of the left posterior medial PhG (TH_L), disappeared in RS_FE, reappeared in RS_1, and was connected with area 4 of the region, which is responsible for the movements of larynx and tongues, and the left PrG (A4tl_L).

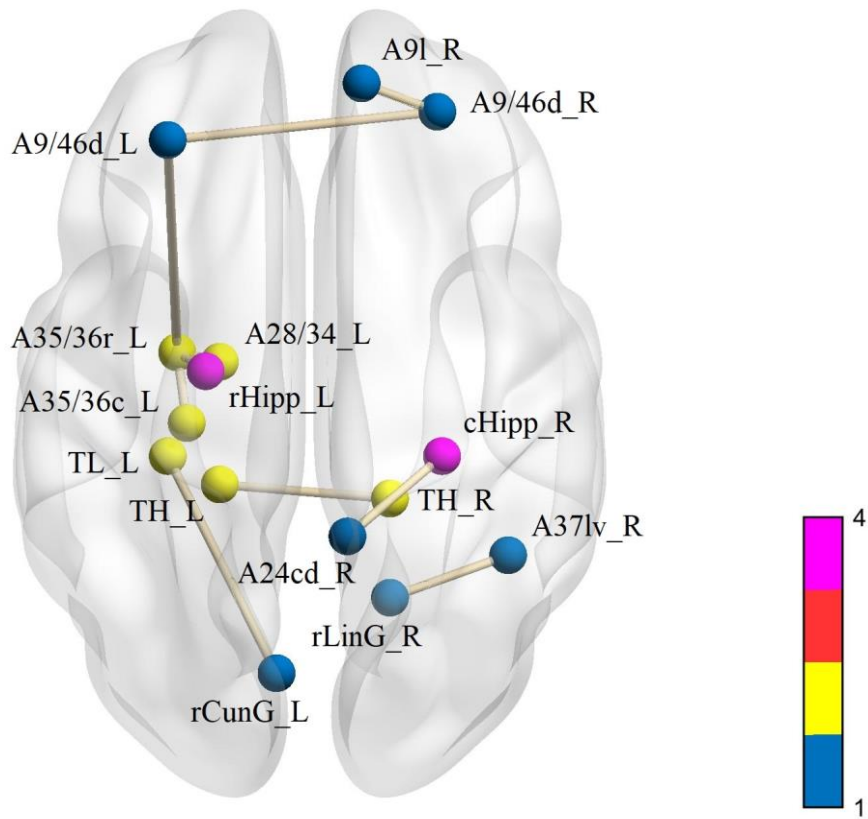

Figure 12. RS_7 – RS_0

Violet – hippocampus

Red – corpus amygdaloideum

Yellow – parahippocampal gyrus

Blue – other areas

In RS_7, a new network, revealed earlier, that consisted of the rostral part of the left PhG (A35/36r_L), the rostral part of the left Hipp (rHipp_L), and the caudal part of the left PhG (A35/36c_L), excluded the left lateral amygdala (lAmyg_L) and re-included area 28/34 of the left entorhinal cortex (A28/34_L) (as in RS_FE). Besides, for the first time, it included the lateral area 9 of the right SFG (A9l_R) and dorsal areas 9/46 of the left and right medium FuG (A9/46d_L, A9/46d_R) (Figure 12). A subnetwork, that consisted of the TH area of the right posterior medial PhG (TH_R), the right rostral lingual gyrus (rLinG_R), and the lateroventral area 37 of the right FuG (A37lv_R), included the TH area of the left posterior medial PhG (TH_L). The caudal part of the right Hipp (cHipp_R) was connected with the ventral area 23 of the right CG (A23v_R). The left posterior lateral PhG (TL_L) was connected with the rostral cuneus of MVOcC (rCunG_R).

DISCUSSION

In the present study, the authors monitored the changes in the brain connectomes of healthy volunteers after the formation and extinction of a fear-conditioned CR within a week after the initial fMRI in the resting condition. The authors used the NBS method to demonstrate the consecutive changes in the brain subnetworks paying special attention to the appearance of new subnetworks that included the regions involved in the emotional training and formation of emotional memory. The authors revealed that new subnetworks that included PhG regions were the most stable. These subnetworks included the areas of Hipp, amygdala, sometimes INS, OrG, temporal and frontal gyri.

Earlier, there were some studies that showed elevated connectivity between amygdalae, temporal, and frontal gyri (LaBar *et al*., 1998; Armony and Dolan**,** 2002; Knight *et al*., 2004; Fullana *et al*., 2016, 2018; Tetereva *et al*., 2020; Martynova *et al*., 2020), as well as PhG (Alvarez e*t al*., 2008), Hipp, INS, and OrG (Greco, 2016; Di *et al*., 2017) during the formation and extinction of a fear-conditioned CR. However, the authors performed the analysis using a more detailed Brainnetome atlas (Fan *et al*., 2016), which allowed them to separate the areas of the ROI and to reveal the effect that persisted for a week.

Thus, the most consecutive development of FC after a fear-conditioned stimulus was observed in a network that appeared around the left dorsal PhG that included the left rostral Hipp in all sessions of the experimental group, the caudal part of the left PhG in the sessions RS_FE, RS_1, RS_7, and left lateral amygdala in the sessions RS_FE and RS_1. The above-mentioned areas are data-driven in the extinction of a fear-conditioned CR (Fullana *et al*., 2016, 2018). The authors also expected to observe elevated connectivity of the amygdala in the session of extinction of a CR (FE) according to the earlier obtained data (Schultz *et al*., 2012; Martynova *et al*., 2020). However, during the extinction time, the appearance of new significant associations was not observed in the amygdalae. In the meta-analysis of the brain activity (not connectivity) performed for fMRI data (Fullana *et al*., 2018) in the procedures of extinction of CR, elevated activity of amygdalae was not observed either, unlike a stable activation of vmPFC, PhG, and Hipp.

In the majority of studies, the researchers separated the Hipp by the dorsoventral axis. However, in the fundamental article (Moser and Moser, 1998), dorsorostral and ventrocaudal division was used according to the generally accepted conception that the dorsal Hipp is closer to the occipital region and rostral – closer to the forehead. In the present study, the authors separated the Hipp by the dorsorostral and ventrocaudal axis. In this case, the obtained results on the appearance of hippocampal subnetwork during and after the extinction of a CR agreed with a generally accepted consensus that the dorsal (caudal) Hipp takes part in the formation and renewal of spatial memory (Goodrich-Hunsaker *et al*., 2008; Keinath *et al*., 2014) and ventral (rostral) Hipp – in the modulation of emotional response (Pohlack *et al*., 2012; Zeidman and Maguire, 2016).

In the present study, the regions of Hipp and PhG showed high coefficients of centrality in the sessions FE, RS_FE, RS_1, which indicated their higher integration in the connectome of the functional brain activity in the resting state observed only after CS+. The main function of PhG is the comparison of data and recognition of a context (Lissek *et al*., 2013; Li *et al*., 2016). Thus, the left rostral PhG modulates the emotional conception of the subject in the association with the areas of the rostral Hipp and amygdala, and the TH area of the posterior medial PhG modulates the spatial context of the experiment together with the caudal Hipp, lingual gyrus, and FuG, which are traditionally considered to be the regions involved in the processing of visual information.

In the session RS_1, the lateral part of the right posterior PhG and the rostral part of the right PhG via the connection with the area 22 of the STG got integrated into a new subnetwork that included the areas of the INS and OrG. At the same time, the left rostral PhG became one of the main connectome hubs. This result is confirmed by a recent study that showed that PhG was a link between the default mode network and the system of memory of temporal gyri (Ward *et al*., 2014).

The identification of subnetworks, present in RS_0 and absent in the consecutive connectomes in the experimental group, showed a certain balance in the connectivity of lateral rostral and caudal areas of the Hipp. Thus, in the session FE, in the connection between the left rostral PhG and the left rostral Hipp, the right rostral PhG loses the connection of the right rostral Hipp. Together with a new connection with the right caudal Hipp in RS_7, the connection of the left caudal Hipp disappears. At the same time, during the search for new subnetworks, new connections were not revealed in the right rostral Hipp or left caudal Hipp. This is not a strict observation but it illustrates the domination of the left rostral Hipp over the right one and right caudal Hipp over the left one. The authors did not find the data on the functional asymmetry of Hipp by the fMRI data, although the anatomic asymmetry of lateral areas of Hipp is a diagnostic sign for certain neurologic diseases (Lucarelli *et al.*, 2012).

In cases with PTSD, earlier, elevated FC in the areas of INS, Hipp (Hughes, 2012), amygdalae, MFG, STG, Tha (Liberzon *et al.*, 1999), and CG (Charney, 2004) in the resting condition was registered. In the present study, their elevated FC was observed in the brain activity in the resting state right after a CS (formation and extinction of a CR after CS+) (RS_FE), a day after CS+ (RS_1), and a week after CS+ (RS_7). The obtained data suggest that after a weak CS+, healthy people have elevated FC of certain areas of the limbic system responsible for emotional training, which later decreases to the initial state approximately a week after CS+. At the same time, in cases with PTSD, the strengthening of elevated connectivity in the regions of the limbic system occurred that was caused by strong CS.

CONCLUSION

The transformed character of the connections in the areas of PhG remained in the brain activity in the default mode for a week after the formation and extinction of a CR. However, the rest areas that showed elevated connectivity in the resting state after CS+ restored the parameters of the connections close to the baseline within a week. At the same time, the parameters of neural networks as a whole-brain connectome restored.

The study was financed by the grant given by the Russian Scientific Foundation No. 16-15-00300.